# Thermodynamic Functions for Body Centered Cubic Lattice- Application on Lattice Green's Function

**J. H. Asad**

Tabuk University, College of Sciences, Department of Physics, P.O.Box.741, Tabuk 7149 1, Kiingdom of Saudi Arabia
Email: jhasad1@yahoo.com.

**Keyword:** Lattice Green's Function (LGF), Body Centered Cubic (BCC) Lattice, and Ionic Systems.

## Abstract

Thermodynamic functions of ionic systems were evaluated analytically using the Green's Function for body centered cubic lattice. The free energy density, chemical potential, pressure, Spinodals, and Coulomb ionic potentials, are expressed in terms of hyper geometric functions $_3F_2$ and complete elliptic integrals.

# I- Introduction

The Lattice Green's Function (LGF) is a basic function in the study of the solid state physics and condensed matter. It appears especially when impure solids are studied [Morita and Horiguchi, 1972]. Green was the first physicist who established the basic concepts of Green's function in the potential theory, and his work was focused on solving Laplace's and Poisson's equations with different boundary conditions. The use of Green's function method plays an important role in many-body problems [Fetter and Walecka, 1971], especially in problems of solid state physics where an enormous progress has been realized. In the mathematical problem of quantum theory which consists of solving linear operator equations with given boundary conditions, Green's functions constitute the natural language to study boundary conditions.

Nowadays, Green's function is one of the most important concepts in many branches of physics, as many quantities in solid state physics can be expressed in terms of LGF. In the following are some examples: statistical model of ferromagnetism such as Ising model [McCoy and Wu, 1978], Heisenberg model [Dalton and Wood, 1967], spherical model [Lax, 1952], random walk theory [Montrol et. al, 1965], [Hughes, 1986], diffusion [Montet, 1973], band structure [Koster and Slater, 1954], Andersion localization in anisotropic systems, such are high Tc superconductors of anisotropic [11,12], resistance calculation for an infinite network of identical resistors [Cserti, 2000], [Cserti et. al, 2002], [Asad 2004], and recently on lattice models of ionic systems [22-29].

The LGF for several structure lattices has been widely studied during the second half of the last century. The first attempts to study the LGF for the Body Centered Cubic (BCC) lattice have been carried out by [Maradudin et al., 1960]. They showed that the LGF for the BCC lattice at the origin $G(0,0,0)$ can be expressed as a product of complete elliptic integrals of the first kind. One can find other useful investigations for the LGF of the BCC lattice in many references as [Joyce, 1971a and b and Inoue, 1975].

In recent years, lattice models have attracted the attention of researchers as a tool for investigating thermodynamics and criticality in Coulomb systems [22,25]. The theoretical studies of lattice models of ionic systems need the LGF For the calculating thermodynamic functions. In this article we will apply the Green's function for a body centered cubic lattice to Debye-Hückel (DH) theory in order to calculate some thermodynamic properties of ionic systems such as total free energy density, chemical potential, pressure, Spinodals, and Coulomb ionic potentials [22,23].

The diagonal Green's function for Body Centered Cubic lattice (BCC) with nearest neighbors interaction is defined by [13,16,18,20].

$$G(0,0,0,E) = \frac{1}{\pi^3} \int_0^\pi \int_0^\pi \int_0^\pi \frac{dk_x dk_y dk_z}{E - Cos(k_x)Cos(k_y)Cos(k_z)} \qquad (1.1)$$

Where $G(0,0,0,E)$ is real for $|E| \geq 1$, and complex for $|E| \leq 1$.

After Solving this integral, and using the analytic continuation, the Green's function at the origin can be written as [16, 20],

$$G(0,0,0;E) = \begin{cases} \dfrac{4}{|E|\pi^2} K^2(k) & |E| > 1 \\ \dfrac{4}{\pi^2} K(k_+)K(k_-) + \dfrac{2i}{\pi^2}[K^2(k_+) - K^2(k_-)] & \end{cases} \qquad (1.2)$$

where K(k) is the complete elliptic integral of the first kind, and

$$k^2 = \frac{1}{2} - \frac{1}{2}\sqrt{1 - E^{-2}}$$

and

$$k_\pm^2 = \frac{1}{2}(1 \pm \sqrt{1 - E^2})$$

## II-Application

The Debye-Hückel (DH) theory developed for dilute solutions of strong electrolytes of Coulomb systems[22-25]. Using the linearized lattice Poisson Boltzman equation, with the DH theory we can write the potential felt by an ion of charge $q_i$ at the origin due to all surrounding ions, electrostatic part of the free energy density, chemical potential, pressure, and Spinodals, (defied by setting the inverse of isothermal compressibility to zero)[22,23], calculating these functions, we will have a full information about the thermodynamics behavior of a lattice Coulomb system. In particular, the possibility of phase transitions and criticality by analyzing the spinodals [22,23].

The Electric potential felt by an ion of charge $q_i$ at the origin due to all surrounding ions which can be derived using linearized lattice Poisson Boltzman equation, with the DH theory as [22,23],

$$\Theta_i = \frac{C_d q_i}{D v_0} \frac{a^2}{6} \left( P(\frac{6}{x^2 + 6}) - P(1) \right) \qquad (2.1)$$

where P(x) is written in terms of LGF as [21]

$$P(\zeta) = \frac{1}{\pi^3} \int_0^\pi \int_0^\pi \int_0^\pi \frac{dk_x dk_y dk_z}{1 - \zeta \cos(k_x)\cos(k_y)\cos(k_z)} = \frac{1}{\zeta} G(0,0,0; \frac{1}{\zeta}) \qquad (2.2)$$

$$C_d = \frac{2\pi^{\frac{3}{2}}}{\Gamma(\frac{3}{2})}$$

where $\Gamma$ is the gamma function, x=ka , $k = \frac{C_d \beta \rho_1 q^2}{D}$ inverse Debye screening length, with $\beta = 1/k_b T$, $v_0 = 4a_0$ volume per site of BCC lattice ($a_0$ is the nearest-neighbor distance),

Using the LGF Eq.(1.2) and Eq. (2.1) , the electric potential for a BCC lattice has the form

$$\Theta_i = \frac{C_d q_i}{D v_0} \frac{a^2}{6} \frac{4}{\pi^2} \left( K^2(\sqrt{\frac{1}{2} - \frac{1}{2}\sqrt{1 - (\frac{6}{x^2+6})^2}}) - K^2(\sqrt{\frac{1}{2}}) \right) \qquad (2.3)$$

The self-potential of an ion (in the absence of any screening) for BCC lattice can be written as [22,23]

$$\phi(0) = \frac{C_d}{v_0} \frac{a^3}{6} (P(1) - 1)$$

$$\phi(0) = \frac{C_d}{v_0} \frac{a^3}{6} \left( \frac{4}{\pi^2} K^2(\sqrt{\frac{1}{2}}) - 1 \right)$$

$$= \frac{1}{2}\sqrt{3}\pi(\frac{\Gamma^4(\frac{1}{4})}{4\pi^3} - 1)$$

=1.069789556517414
≈1.070 Kobelev's results [22].

Now the electrostatic part of the Helmholtz free energy which is very important for the determination of the thermodynamic behavior of the ionic systems, can be written as [22]

$$\bar{f}^{EL} = \frac{1}{4dv_0}\left(x^2 P(1) - \int_0^{x^2} P(\frac{6}{x^2+6})d(x^2)\right) \qquad (2.4)$$

Using the LGF Eq.(1.2) and Eq. (2.4), the free energy for a BCC lattice has the form

$$\bar{f}^{EL} = \frac{1}{4dv_0}\frac{4}{\pi^2}\left(x^2 K^2(\sqrt{\frac{1}{2}}) - \int_0^{x^2} K^2(\sqrt{\frac{1}{2} - \frac{1}{2}\sqrt{1-(\frac{6}{x^2+6})^2}})d(x^2)\right) \qquad (2.5)$$

Eq. (2.5) can be integrated, as in appendix A, the free energy for a BCC lattice can be written in a closed form, in terms of elliptic integrals, and generalized hypergeometric function $_3F_2$ as,

$$\bar{f}^{EL} = \frac{1}{4dv_0}\left(\frac{4}{\pi^2}x^2 K^2(\sqrt{\frac{1}{2}}) - [-6\,_3F_2(-1/2,1/2,1/2;1,1;1) + (6+x^2)\,_3F_2((-1/2,1/2,1/2;1,1;\frac{36}{(6+x^2)^2})]\right) \qquad (2.6)$$

where $_3F_2$ is the generalized hypergeometric function.

The chemical potential is defined as
$$\bar{\mu} = \frac{\mu}{k_B T} = -\frac{\partial \bar{f}}{\partial \rho}$$

By differentiate the free energy with respect to density, one can write the the electrostatic part of the chemical potential for each type of ion as [22]

$$\bar{\mu}_1 = -\frac{C_d}{12v_0}\frac{a^3}{T^*}\left(P(1) - P(\frac{6}{x^2+6})\right) \qquad (2.7)$$

where $T^*$ is the reduced temperature and defined by

$$T^* = \frac{k_B TDa}{q^2}$$

Using the LGF Eq.(1.2) and Eq. (2.7), the chemical potential for a BCC lattice has the form

$$\bar{\mu}_1 = -\frac{C_d}{12v_0} \frac{a^3}{T^*} \left( \left( K^2(\sqrt{\frac{1}{2}}) - K^2(\sqrt{\frac{1}{2} - \frac{1}{2}\sqrt{1-(\frac{6}{x^2+6})^2}}) \right) \right) \qquad (2.8)$$

from the chemical potential and the free energy, we can write the electrostatic part of the pressure for a BCC lattice as [22,23]

$$\bar{p}v_0 = \frac{1}{12} \left( x^2 P(\frac{6}{x^2+6}) - \int_0^{x^2} P(\frac{6}{x^2+6}) d(x^2) \right) \qquad (2.9)$$

Using the LGF Eq.(1.2) and Eq. (2.9), the pressure for a BCC lattice can be written as

$$\bar{p}v_0 = \frac{1}{12} \frac{4}{\pi^2} \left( x^2 K^2(\sqrt{\frac{1}{2} - \frac{1}{2}\sqrt{1-(\frac{6}{x^2+6})^2}}) - \int_0^{x^2} K^2(\sqrt{\frac{1}{2} - \frac{1}{2}\sqrt{1-(\frac{6}{x^2+6})^2}}) d(x^2) \right) \qquad (2.10)$$

Eq. (2.10) can be integrated, as in appendix A, the pressure for a BCC lattice can be written in a closed form, in terms of elliptic integrals, and generalized hypergeometric function $_3F_2$ as,

$$\bar{p}v_0 = \frac{1}{12} \left( \begin{array}{l} x^2 \,_3F_2(1/2,1/2,1/2;1,1;(\frac{6}{x^2+6})^2) - [-6 \,_3F_2(-1/2,1/2,1/2;1,1;1) + \\ (6+x^2) \,_3F_2((-1/2,1/2,1/2;1,1;\frac{36}{(6+x^2)^2})] \end{array} \right) \qquad (2.11)$$

the thermodynamic description of the BCC lattice at the critical region can be investigated by using the Spinodals which is defined by $\rho(\partial \bar{\mu}/\partial \rho) = 0$ [22,23], using the chemical potential Eq.(2.7), the Spinodals can be written as

$$T_s^* = \frac{C_3 a^3}{6v_0} \frac{u(1-u)\frac{\partial P(u)}{\partial u}}{2+(1-u)^2 \frac{\partial P(u)}{\partial u}} \qquad (2.12)$$

with
$$u = \frac{6}{x^2 + 6}$$

Using the LGF Eq.(1.2) and Eq. (2.12), the spinodal for a BCC lattice has the form

$$T_s^* = \frac{C_3 a^3}{6v_0} \frac{u(1-u)\frac{4}{\pi^2}K(k)(\frac{E(k)-(1-k^2)K(k)}{k(1-k^2)})(\frac{-u}{k\sqrt{1-u^2}})}{2+(1-u)^2\frac{4}{\pi^2}K(k)(\frac{E(k)-(1-k^2)K(k)}{k(1-k^2)})(\frac{-u}{k\sqrt{1-u^2}})} \quad (2.13)$$

Now the full thermodynamic description for the BCC lattice can be investigated using Eqs.(2.3, 2.6, 2.8, 2.11, 2.13).

### III- Results and Discussion

We have investigated the Green's function for the body centered cubic lattice and derive some thermodynamic functions analytically which are useful in DH theory with Bjerrum clustering, cluster-ion interactions, phase transitions study, and electrolyte systems analysis[22,23].

We see from Figs. (1-5) the graphs of the thermodynamic functions, which are useful to understand the behavior of BCC lattice thermodynamically.

Fig. 1 shows the gas-liquid coexistence curve predicted by pure DH theory, and from the graph we see that the maximum value occurs at x=0.978965 and $T^*$ =0.0369823, which specifies the critical point for BCC lattice. Fig. (2) shows the behavior of the of the potential felt by an ion as a function of x (x=ka), we see that when x becomes large the potential becomes small, and The self-potential of an ion (in the absence of any screening) for BCC lattice is 1.069789556517414 . Fig(3) shows the behavior of free energy versus x. Fig. (4) give us an idea about the behavior of chemical potential versus x. Fig. (5) Shows the behavior of the pressure.

# Figure Captions

**Fig. 1:** Predicted phase diagram of gas-liquid coexistence for BCC lattice ($T^*$ versus x) at arbitrary units

**Fig. 2:** The behavior of the potential felt by an ion as a function of x at arbitrary units.

**Fig. 3:** The behavior of free energy versus x, at arbitrary units.

**Fig. 4:** The behavior of chemical potential versus x, at arbitrary units.

**Fig. 5:** The behavior of the pressure versus x, at arbitrary units.

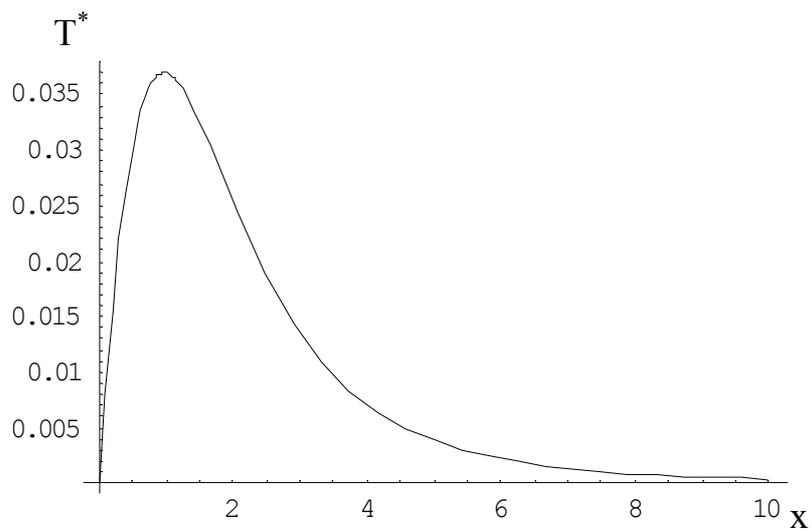

Fig. 1

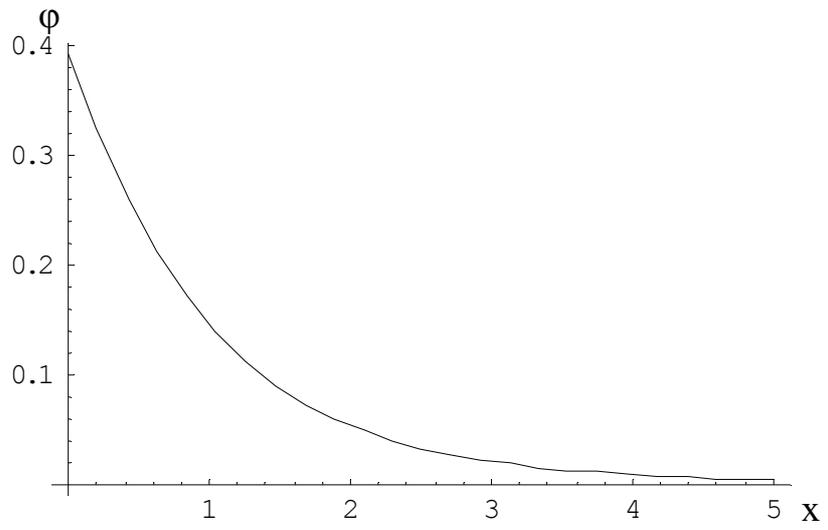

Fig. 2

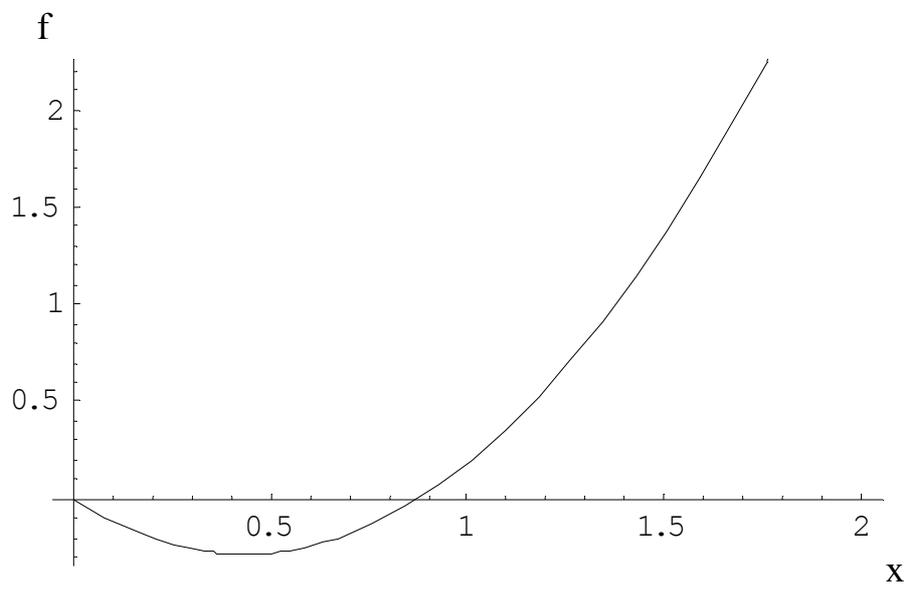

Fig. 3

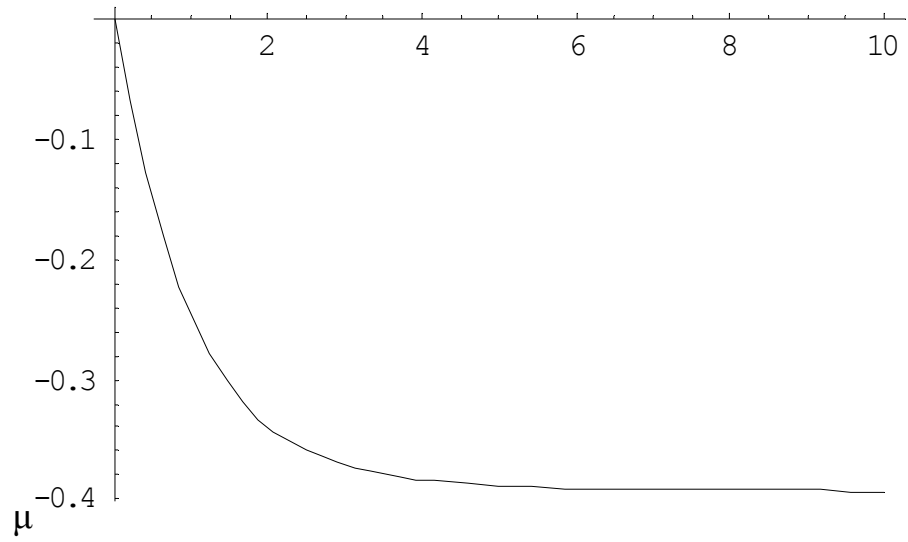

Fig. 4

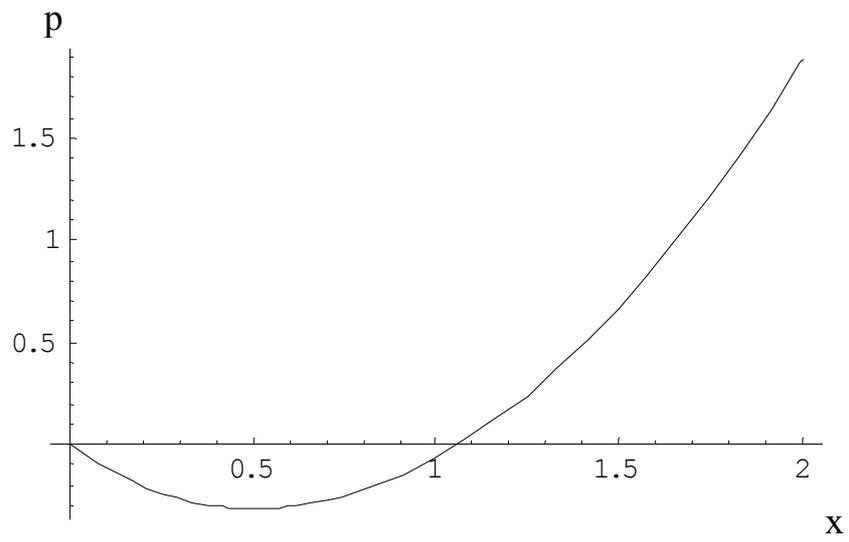

Fig. 5

**References:**

1- P. H. Dederichs, K. Shroeder, and R. Zeller, " Point Defect in Metals II" , (Springer- Verlag, Berlin 1980).

2- E. W. Montroll, and G. H. Weiss, J. Math. Phys., **6**, 2, 167 (1965)

3- G. N. Watson, Quart. J. Math. , (Oxford), **10**, 266 (1939).

4- E. N. Economou, " Green's Functions in Quantum Physics", 2ed edition , (Springer- Verlag, Berlin 1983).

5- R. Brout, Phys. Rev., **118**,1009 , (1960).

6- N. W. Dalton, and D. W. Wood, Proc. Phys. Soc. (London), **90**,4591, (1967).

7- M. Lax, Phys. Rev. , **97**, 629, (1955).

8- E. Montroll and G. Weiss, J. Math. Phys., **6**, 168, (1965).

9- B. Hughes, J. Math. Phys., **23**,1688, (1982).

10- G. F. Koster and J. C. Slater, Phys. Rev., **94**, 1498, (1954).

11- V. V. Bryksin, and P. Kleinert, Z. Phys. B (Condensed matter), 168, (1993).

12- G. L. Montet, Phys. Rev. B, **7**, 650, (1973).

13- V.K. Tewary, adv.in Phys. 757 (1973).

14- Q. Li, C. Soukoulis, E. N. Economou, and G. S. Grest, Phys. Rev. B, **20**, 2825,(1989-I).

15- A. Erdelyi, W. Magnus, F. Oberhettinger, and F. Tricomi, Bateman Manuscript Project ," Higher Transcendential Functions (3 Vol), (McGraw-Hill, New York 1953).

16- G. S. Joyce. J. Math. Phys., 12, 7 (1971).

17- J. Cserti, Am. J. Phys., **68**, 896, (2000).

18- S. Katsura, and T. Horiguchi, J. Math. Phys. 12, 2 (1971)

19- T. Morita, and T. Horiguchi, J. Phys. A. Gen. Phys. , **5**, 67, (1972).

20- A . Sakaji, R. Hijjawi, N. Shawagfeh, and J. Kalifeh, Inte. J. Theo. Phys., **41**, 5, 973 (2002). J. Math. Phys. **43**, 1, 235 (2002).

21- G. S. Joyce , J. Phys. A: Math. Gen. , **36**, 911, (2003).

22 -V. Kobelev, and A. Kolomeisky, J. Chem. Phys. **116**, 17, 7589, (2002).

23- V. Kobelev, and A. Kolomeisky, J. Chem. Phys. **116**, 19, 8879, (2002).

24- M. E. Fisher, J. Stat. Phys. **76**, 1 (1994).

25- M. E. Fisher, J. Phys. Cond. Matt. **8**, 9103 (1996).

26- G. Stell, J. Stat. Phys. **78**, 197 (1995).

27- G. Stell, J. Cond. Matt. **8**, 9329 (1996).

28- H. Weingartner and W. Schroer, Adv. Chem. Phys. **116**,1 (2001).

29- Y. Levin, Repts. Progr. Phys. **65**, 1577 (2002).

30- Qimiao Si, S. Rabello, K. Ingersent, and L. Smith, Nature, 413, 804, (2001).

# Appendix A

Consider the following integral

$$I = \int_0^{x^2} P(\frac{6}{x^2+6}) d(x^2) \qquad (A1)$$

using the LGF Eqs.( 1.2, 2.2), and expand the elliptic integral in terms of the generalized hypergeometric function , we can write A1 as

$$I = \int_0^{x^2} {}_3F_2(1/2,1/2,1/2;1,1;(\frac{6}{x^2+6})^2) d(x^2) \qquad (A2)$$

by direct substitution , $s=x^2$ Eq. (A2) can be written as

$$I = \int_0^s {}_3F_2(1/2,1/2,1/2;1,1;(\frac{6}{s+6})^2) d(s) \qquad (A3)$$

Since this function can be expanded as a uniform convergence series, by expanding the hypergeometric function in Eq. (A3) and integrate term by term we have,

$$I = -6 {}_3F_2(-1/2,1/2,1/2;1,1;1) + (6+s) {}_3F_2(-1/2,1/2,1/2;1,1;(\frac{36}{(6+s)^2})) \qquad (A4)$$